
\magnification=\magstephalf
\def\Title#1{%
\vskip 1in{\titlefont\centerline{#1}}\vskip .5in}

\def\Date#1{\leftline{#1}\tenrm\supereject%
\global\hsize=\hsbody\global\hoffset=\hbodyoffset%
\footline={\hss\tenrm\folio\hss}}

\newif\ifdraftmode
\newif\ifleftlabels  

\def\nolabels{\def\wrlabeL##1{}\def\eqlabeL##1{}\def\reflabeL##1{}}
\def\writelabels{\def\wrlabeL##1{\leavevmode\vadjust{\rlap{\smash%
{\line{{\escapechar=` \hfill\rlap{\sevenrm\hskip.03in\string##1}}}}}}}%
\def\eqlabeL##1{{\escapechar-1\rlap{\sevenrm\hskip.05in\string##1}}}%
\def\reflabeL##1{\noexpand\rlap{\noexpand\sevenrm[\string##1]}}}
\def\writeleftlabels{\def\wrlabeL##1{\leavevmode\vadjust{\rlap{\smash%
{\line{{\escapechar=` \hfill\rlap{\sevenrm\hskip.03in\string##1}}}}}}}%
\def\eqlabeL##1{{\escapechar-1%
\rlap{\sixrm\hskip.05in\string##1}%
\llap{\sevenrm\string##1\hskip.03in\hbox to \hsize{}}}}%
\def\reflabeL##1{\noexpand\rlap{\noexpand\sevenrm[\string##1]}}}
\nolabels

\newdimen\fullhsize
\newdimen\hstitle
\hstitle=\hsize 
\newdimen\hsbody
\hsbody=\hsize 
\newdimen\hbodyoffset
\hbodyoffset=\hoffset 
\newbox\leftpage
\def\abstract#1{#1}
\def\rotated{\special{ps: landscape}
\magnification=1000  
\baselineskip=14pt
\global\hstitle=9truein\global\hsbody=4.75truein
\global\vsize=7truein\global\voffset=-.31truein
\global\hoffset=-0.54in\global\hbodyoffset=-.54truein
\global\fullhsize=10truein
\def\DefineTeXgraphics{%
\special{ps::[global]
/TeXgraphics {currentpoint translate 0.7 0.7 scale
              -80 0.72 mul -1000 0.72 mul translate} def}}
\let\lr=L
\def\ifsmall{\iftrue}
\def\titlepagefont{\twelvepoint}
\trueseventeenpoint
\def\almostshipout##1{\if L\lr \count1=1
      \global\setbox\leftpage=##1 \global\let\lr=R
   \else \count1=2
      \shipout\vbox{\hbox to\fullhsize{\box\leftpage\hfil##1}}
      \global\let\lr=L\fi}

\output={\ifnum\count0=1 
 \shipout\vbox{\hbox to \fullhsize{\hfill\pagebody\hfill}}\advancepageno
 \else
 \almostshipout{\leftline{\vbox{\pagebody\makefootline}}}\advancepageno
 \fi}

\def\abstract##1{{\leftskip=1.5in\rightskip=1.5in ##1\par}} }

\def\linemessage#1{\immediate\write16{#1}}

\global\newcount\secno \global\secno=0
\global\newcount\appno \global\appno=0
\global\newcount\meqno \global\meqno=1
\global\newcount\subsecno \global\subsecno=0
\global\newcount\figno \global\figno=0

\newif\ifAnyCounterChanged
\let\terminator=\relax
\def\normalize#1{\ifx#1\terminator\let\next=\relax\else%
\if#1i\aftergroup i\else\if#1v\aftergroup v\else\if#1x\aftergroup x%
\else\if#1l\aftergroup l\else\if#1c\aftergroup c\else%
\if#1m\aftergroup m\else%
\if#1I\aftergroup I\else\if#1V\aftergroup V\else\if#1X\aftergroup X%
\else\if#1L\aftergroup L\else\if#1C\aftergroup C\else%
\if#1M\aftergroup M\else\aftergroup#1\fi\fi\fi\fi\fi\fi\fi\fi\fi\fi\fi\fi%
\let\next=\normalize\fi%
\next}
\def\makeNormal#1#2{\def\doNormalDef{\edef#1}\begingroup%
\aftergroup\doNormalDef\aftergroup{\normalize#2\terminator\aftergroup}%
\endgroup}

\def\warnIfChanged#1#2{%
\ifundef#1
\else\begingroup%
\edef\oldDefinitionOfCounter{#1}\edef\newDefinitionOfCounter{#2}%
\ifx\oldDefinitionOfCounter\newDefinitionOfCounter%
\else%
\linemessage{Warning: definition of \noexpand#1 has changed.}%
\global\AnyCounterChangedtrue\fi\endgroup\fi}

\def\Section#1{\global\advance\secno by1\relax\global\meqno=1%
\global\subsecno=0%
\bigbreak\bigskip
\centerline{\twelvepoint \bf %
\the\secno. #1}%
\par\nobreak\medskip\nobreak}
\def\tagsection#1{%
\warnIfChanged#1{\the\secno}%
\xdef#1{\the\secno}%
\ifWritingAuxFile\immediate\write\auxfile{\noexpand\xdef\noexpand#1{#1}}\fi%
}
\def\section{\Section}
\def\Subsection#1{\global\advance\subsecno by1\relax\medskip %
\leftline{\bf\the\secno.\the\subsecno\ #1}%
\par\nobreak\smallskip\nobreak}
\def\tagsubsection#1{%
\warnIfChanged#1{\the\secno.\the\subsecno}%
\xdef#1{\the\secno.\the\subsecno}%
\ifWritingAuxFile\immediate\write\auxfile{\noexpand\xdef\noexpand#1{#1}}\fi%
}

\def\subsection{\Subsection}

\def\romappno{\uppercase\expandafter{\romannumeral\appno}}
\def\makeNormalizedRomappno{%
\expandafter\makeNormal\expandafter\normalizedromappno%
\expandafter{\romannumeral\appno}%
\edef\normalizedromappno{\uppercase{\normalizedromappno}}}
\def\Appendix#1{\global\advance\appno by1\relax\global\meqno=1\global\secno=0%
\global\subsecno=0%
\bigbreak\bigskip
\centerline{\twelvepoint \bf Appendix %
\romappno. #1}%
\par\nobreak\medskip\nobreak}
\def\tagappendix#1{\makeNormalizedRomappno%
\warnIfChanged#1{\normalizedromappno}%
\xdef#1{\normalizedromappno}%
\ifWritingAuxFile\immediate\write\auxfile{\noexpand\xdef\noexpand#1{#1}}\fi%
}
\def\appendix{\Appendix}
\def\Subappendix#1{\global\advance\subsecno by1\relax\medskip %
\leftline{\bf\romappno.\the\subsecno\ #1}%
\par\nobreak\smallskip\nobreak}
\def\tagsubappendix#1{\makeNormalizedRomappno%
\warnIfChanged#1{\normalizedromappno.\the\subsecno}%
\xdef#1{\normalizedromappno.\the\subsecno}%
\ifWritingAuxFile\immediate\write\auxfile{\noexpand\xdef\noexpand#1{#1}}\fi%
}

\def\eqn#1{\makeNormalizedRomappno%
\ifnum\secno>0%
  \warnIfChanged#1{\the\secno.\the\meqno}%
  \eqno(\the\secno.\the\meqno)\xdef#1{\the\secno.\the\meqno}%
     \global\advance\meqno by1
\else\ifnum\appno>0%
  \warnIfChanged#1{\normalizedromappno.\the\meqno}%
  \eqno({\rm\romappno}.\the\meqno)%
      \xdef#1{\normalizedromappno.\the\meqno}%
     \global\advance\meqno by1
\else%
  \warnIfChanged#1{\the\meqno}%
  \eqno(\the\meqno)\xdef#1{\the\meqno}%
     \global\advance\meqno by1
\fi\fi%
\eqlabeL#1%
\ifWritingAuxFile\immediate\write\auxfile{\noexpand\xdef\noexpand#1{#1}}\fi%
}
\def\defeqn#1{\makeNormalizedRomappno%
\ifnum\secno>0%
  \warnIfChanged#1{\the\secno.\the\meqno}%
  \xdef#1{\the\secno.\the\meqno}%
     \global\advance\meqno by1
\else\ifnum\appno>0%
  \warnIfChanged#1{\normalizedromappno.\the\meqno}%
  \xdef#1{\normalizedromappno.\the\meqno}%
     \global\advance\meqno by1
\else%
  \warnIfChanged#1{\the\meqno}%
  \xdef#1{\the\meqno}%
     \global\advance\meqno by1
\fi\fi%
\eqlabeL#1%
\ifWritingAuxFile\immediate\write\auxfile{\noexpand\xdef\noexpand#1{#1}}\fi%
}
\def\anoneqn{\makeNormalizedRomappno%
\ifnum\secno>0
  \eqno(\the\secno.\the\meqno)%
     \global\advance\meqno by1
\else\ifnum\appno>0
  \eqno({\rm\normalizedromappno}.\the\meqno)%
     \global\advance\meqno by1
\else
  \eqno(\the\meqno)%
     \global\advance\meqno by1
\fi\fi%
}
\def\mfig#1#2{\global\advance\figno by1%
\relax#1\the\figno%
\warnIfChanged#2{\the\figno}%
\edef#2{\the\figno}%
\reflabeL#2%
\ifWritingAuxFile\immediate\write\auxfile{\noexpand\xdef\noexpand#2{#2}}\fi%
}

\catcode`@=11 

\font\ninerm=cmr9
\font\eightrm=cmr8
\font\sixrm=cmr6

\def\loadtrueseventeenpoint{
 \font\seventeenrm=cmr10 at 17.28truept
 \font\seventeeni=cmmi10 at 17.28truept
 \font\seventeenbf=cmbx10 at 17.28truept
 \font\seventeenit=cmti10 at 17.28truept
 \font\seventeensl=cmsl10 at 17.28truept
 \font\seventeensy=cmsy10 at 17.28truept
}
\def\loadfourteenpoint{
\font\fourteenrm=cmr10 at 14.4pt
\font\fourteeni=cmmi10 at 14.4pt
\font\fourteenit=cmti10 at 14.4pt
\font\fourteensl=cmsl10 at 14.4pt
\font\fourteensy=cmsy10 at 14.4pt
\font\fourteenbf=cmbx10 at 14.4pt
}
\def\loadtruetwelvepoint{
\font\twelverm=cmr10 at 12truept
\font\twelvei=cmmi10 at 12truept
\font\twelveit=cmti10 at 12truept
\font\twelvesl=cmsl10 at 12truept
\font\twelvesy=cmsy10 at 12truept
\font\twelvebf=cmbx10 at 12truept
}

\font\ninei=cmmi9
\font\eighti=cmmi8
\font\sixi=cmmi6
\skewchar\ninei='177 \skewchar\eighti='177 \skewchar\sixi='177

\font\ninesy=cmsy9
\font\eightsy=cmsy8
\font\sixsy=cmsy6
\skewchar\ninesy='60 \skewchar\eightsy='60 \skewchar\sixsy='60

\font\ninebf=cmbx9
\font\eightbf=cmbx10
\font\sixbf=cmbx6

\font\ninett=cmtt9
\font\eighttt=cmtt8

\hyphenchar\tentt=-1 
\hyphenchar\ninett=-1
\hyphenchar\eighttt=-1

\font\ninesl=cmsl9
\font\eightsl=cmsl8

\font\nineit=cmti9
\font\eightit=cmti8


\newskip\ttglue
\def\tenpoint{\def\rm{\fam0\tenrm}%
  \textfont0=\tenrm \scriptfont0=\sevenrm \scriptscriptfont0=\fiverm
  \textfont1=\teni \scriptfont1=\seveni \scriptscriptfont1=\fivei
  \textfont2=\tensy \scriptfont2=\sevensy \scriptscriptfont2=\fivesy
  \textfont3=\tenex \scriptfont3=\tenex \scriptscriptfont3=\tenex
  \def\it{\fam\itfam\tenit}\textfont\itfam=\tenit
  \def\sl{\fam\slfam\tensl}\textfont\slfam=\tensl
  \def\bf{\fam\bffam\tenbf}\textfont\bffam=\tenbf \scriptfont\bffam=\sevenbf
  \scriptscriptfont\bffam=\fivebf
  \normalbaselineskip=12pt
  \let\sc=\eightrm
  \let\big=\tenbig
  \setbox\strutbox=\hbox{\vrule height8.5pt depth3.5pt width\z@}%
  \normalbaselines\rm}

\def\twelvepoint{\def\rm{\fam0\twelverm}%
  \textfont0=\twelverm \scriptfont0=\ninerm \scriptscriptfont0=\sevenrm
  \textfont1=\twelvei \scriptfont1=\ninei \scriptscriptfont1=\seveni
  \textfont2=\twelvesy \scriptfont2=\ninesy \scriptscriptfont2=\sevensy
  \textfont3=\tenex \scriptfont3=\tenex \scriptscriptfont3=\tenex
  \def\it{\fam\itfam\twelveit}\textfont\itfam=\twelveit
  \def\sl{\fam\slfam\twelvesl}\textfont\slfam=\twelvesl
  \def\bf{\fam\bffam\twelvebf}\textfont\bffam=\twelvebf%
  \scriptfont\bffam=\ninebf
  \scriptscriptfont\bffam=\sevenbf
  \normalbaselineskip=12pt
  \let\sc=\eightrm
  \let\big=\tenbig
  \setbox\strutbox=\hbox{\vrule height8.5pt depth3.5pt width\z@}%
  \normalbaselines\rm}

\def\fourteenpoint{\def\rm{\fam0\fourteenrm}%
  \textfont0=\fourteenrm \scriptfont0=\tenrm \scriptscriptfont0=\sevenrm
  \textfont1=\fourteeni \scriptfont1=\teni \scriptscriptfont1=\seveni
  \textfont2=\fourteensy \scriptfont2=\tensy \scriptscriptfont2=\sevensy
  \textfont3=\tenex \scriptfont3=\tenex \scriptscriptfont3=\tenex
  \def\it{\fam\itfam\fourteenit}\textfont\itfam=\fourteenit
  \def\sl{\fam\slfam\fourteensl}\textfont\slfam=\fourteensl
  \def\bf{\fam\bffam\fourteenbf}\textfont\bffam=\fourteenbf%
  \scriptfont\bffam=\tenbf
  \scriptscriptfont\bffam=\sevenbf
  \normalbaselineskip=17pt
  \let\sc=\elevenrm
  \let\big=\tenbig
  \setbox\strutbox=\hbox{\vrule height8.5pt depth3.5pt width\z@}%
  \normalbaselines\rm}

\def\seventeenpoint{\def\rm{\fam0\seventeenrm}%
  \textfont0=\seventeenrm \scriptfont0=\fourteenrm \scriptscriptfont0=\tenrm
  \textfont1=\seventeeni \scriptfont1=\fourteeni \scriptscriptfont1=\teni
  \textfont2=\seventeensy \scriptfont2=\fourteensy \scriptscriptfont2=\tensy
  \textfont3=\tenex \scriptfont3=\tenex \scriptscriptfont3=\tenex
  \def\it{\fam\itfam\seventeenit}\textfont\itfam=\seventeenit
  \def\sl{\fam\slfam\seventeensl}\textfont\slfam=\seventeensl
  \def\bf{\fam\bffam\seventeenbf}\textfont\bffam=\seventeenbf%
  \scriptfont\bffam=\fourteenbf
  \scriptscriptfont\bffam=\twelvebf
  \normalbaselineskip=21pt
  \let\sc=\fourteenrm
  \let\big=\tenbig
  \setbox\strutbox=\hbox{\vrule height 12pt depth 6pt width\z@}%
  \normalbaselines\rm}

\def\ninepoint{\def\rm{\fam0\ninerm}%
  \textfont0=\ninerm \scriptfont0=\sixrm \scriptscriptfont0=\fiverm
  \textfont1=\ninei \scriptfont1=\sixi \scriptscriptfont1=\fivei
  \textfont2=\ninesy \scriptfont2=\sixsy \scriptscriptfont2=\fivesy
  \textfont3=\tenex \scriptfont3=\tenex \scriptscriptfont3=\tenex
  \def\it{\fam\itfam\nineit}\textfont\itfam=\nineit
  \def\sl{\fam\slfam\ninesl}\textfont\slfam=\ninesl
  \def\bf{\fam\bffam\ninebf}\textfont\bffam=\ninebf \scriptfont\bffam=\sixbf
  \scriptscriptfont\bffam=\fivebf
  \normalbaselineskip=11pt
  \let\sc=\sevenrm
  \let\big=\ninebig
  \setbox\strutbox=\hbox{\vrule height8pt depth3pt width\z@}%
  \normalbaselines\rm}

\def\eightpoint{\def\rm{\fam0\eightrm}%
  \textfont0=\eightrm \scriptfont0=\sixrm \scriptscriptfont0=\fiverm%
  \textfont1=\eighti \scriptfont1=\sixi \scriptscriptfont1=\fivei%
  \textfont2=\eightsy \scriptfont2=\sixsy \scriptscriptfont2=\fivesy%
  \textfont3=\tenex \scriptfont3=\tenex \scriptscriptfont3=\tenex%
  \def\it{\fam\itfam\eightit}\textfont\itfam=\eightit%
  \def\sl{\fam\slfam\eightsl}\textfont\slfam=\eightsl%
  \def\bf{\fam\bffam\eightbf}\textfont\bffam=\eightbf \scriptfont\bffam=\sixbf%
  \scriptscriptfont\bffam=\fivebf%
  \normalbaselineskip=9pt%
  \let\sc=\sixrm%
  \let\big=\eightbig%
  \setbox\strutbox=\hbox{\vrule height7pt depth2pt width\z@}%
  \normalbaselines\rm}

\def\tenbig#1{{\hbox{$\left#1\vbox to8.5pt{}\right.\n@space$}}}
\def\ninebig#1{{\hbox{$\textfont0=\tenrm\textfont2=\tensy
  \left#1\vbox to7.25pt{}\right.\n@space$}}}
\def\eightbig#1{{\hbox{$\textfont0=\ninerm\textfont2=\ninesy
  \left#1\vbox to6.5pt{}\right.\n@space$}}}

\def\footnote#1{\edef\@sf{\spacefactor\the\spacefactor}#1\@sf
      \insert\footins\bgroup\eightpoint
      \interlinepenalty100 \let\par=\endgraf
        \leftskip=\z@skip \rightskip=\z@skip
        \splittopskip=10pt plus 1pt minus 1pt \floatingpenalty=20000
        \smallskip\item{#1}\bgroup\strut\aftergroup\@foot\let\next}
\skip\footins=12pt plus 2pt minus 4pt 
\dimen\footins=30pc 

\newinsert\margin
\dimen\margin=\maxdimen
\def\titlefont{\seventeenpoint}
\loadtruetwelvepoint 
\loadtrueseventeenpoint

\def\eatOne#1{}
\def\ifundef#1{\expandafter\ifx%
\csname\expandafter\eatOne\string#1\endcsname\relax}
\def\notTrue{\iffalse}\def\isTrue{\iftrue}
\def\ifdef#1{{\ifundef#1%
\aftergroup\notTrue\else\aftergroup\isTrue\fi}}
\def\use#1{\ifundef#1\linemessage{Warning: \string#1 is undefined.}%
{\tt \string#1}\else#1\fi}


\global\newcount\refno \global\refno=1
\newwrite\rfile
\newlinechar=`\^^J
\def\@ref#1#2{\the\refno\n@ref#1{#2}}
\def\n@ref#1#2{\xdef#1{\the\refno}%
\ifnum\refno=1\immediate\openout\rfile=\jobname.refs\fi%
\immediate\write\rfile{\noexpand\item{[\noexpand#1]\ }#2.}%
\global\advance\refno by1}
\def\nref{\n@ref} 
\def\ref{\@ref}   
\def\lref#1#2{\the\refno\xdef#1{\the\refno}%
\ifnum\refno=1\immediate\openout\rfile=\jobname.refs\fi%
\immediate\write\rfile{\noexpand\item{[\noexpand#1]\ }#2\semi}%
\global\advance\refno by1}
\def\cref#1{\immediate\write\rfile{#1\semi}}

\def\preref#1#2{\gdef#1{\@ref#1{#2}}}

\def\semi{;\hfil\noexpand\break}

\def\listrefs{\vfill\eject\immediate\closeout\rfile
\centerline{{\bf References}}\bigskip\frenchspacing%
\input \jobname.refs\vfill\eject\nonfrenchspacing}

\def\inputAuxIfPresent#1{\immediate\openin1=#1
\ifeof1\message{No file \auxfileName; I'll create one.
}\else\closein1\relax\input\auxfileName\fi%
}

\newif\ifWritingAuxFile
\newwrite\auxfile
\def\SetUpAuxFile{%
\xdef\auxfileName{\jobname.aux}%
\inputAuxIfPresent{\auxfileName}%
\WritingAuxFiletrue%
\immediate\openout\auxfile=\auxfileName}



\catcode`\@=\active
\catcode`@=12  
\catcode`\"=\active

\SetUpAuxFile
\loadfourteenpoint

\nopagenumbers\hsize=\hstitle\vskip1in
\overfullrule 0pt
\hfuzz 35 pt
\vbadness=10001
%
%
\def\"#1{{\accent127 #1}}

\noindent
hep-ph/9504424 \hfil \break
\rightline{PITHA 95/11}
\rightline{IKDA-95/16}
\rightline{April, 1995}
\rightline{   }

\leftlabelstrue
\vskip -1.0 in
\Title{ Probing
Higgs Sector CP Violation
with Top Quarks}
\vskip-3.8cm
\Title{
at a Photon Linear Collider}
\centerline{Harald Anlauf${}^{\dagger}$}
\baselineskip12truept
\centerline{\it Institut f\"ur  Kernphysik, TH Darmstadt}
\centerline{\it 64289 Darmstadt, Germany }
\centerline{\it and}
\centerline{\it  Fachbereich Physik, Universit\"at Siegen}
\centerline{\it  57076 Siegen, Germany}

\smallskip\smallskip

\baselineskip17truept
\centerline{Werner Bernreuther and
 Arnd Brandenburg}
\baselineskip12truept
\centerline{\it Institut f\"ur Theoretische Physik}
\centerline{\it Physikzentrum}
\centerline{\it Rheinisch-Westf\"alische Technische Hochschule Aachen}
\centerline{\it 52056 Aachen, Germany}

\vskip 0.2in\baselineskip13truept

\vskip 0.5truein
\centerline{\bf Abstract}

{\narrower
We investigate for two-Higgs doublet models with intrinsic CP violation in
the scalar potential
CP-nonconserving effects in unpolarized
  $\gamma \gamma \rightarrow t \bar{t} $ for a range of
neutral Higgs boson masses which includes
 resonant $\varphi$ production and the subsequent
decay of $\varphi \rightarrow t \bar{t}$.
The importance of taking into account, even in the resonant case,
the  interference with the
nonresonant background is shown.
Further, we propose and calculate
three  asymmetries which efficiently trace CP-violating effects in
$\gamma \gamma \rightarrow t \bar{t} $ using
semileptonic $t \bar{t}$ decays.}

\vskip 0.3truein

\

\vfill
\vskip 0.1in
\noindent\hrule width 3.6in\hfil\break

\noindent
${}^{\dagger}$Research supported in part by the Deutsche
Forschungsgemeinschaft.
\hfil\break

\Date{}

\line{}

\baselineskip17pt
%


\preref\Cohen{%
A.G. Cohen, D.B. Kaplan, and A.E. Nelson,
Ann.\ Rev.\ Nucl.\ Part.\ Sci.\ {\bf 43} (1993) 27}

\preref\Lee{%
T.D. Lee, Phys.\ Rev.\ {\bf D8} (1973); J. Liu and L.
Wolfenstein, Nucl.\ Phys.\ {\bf B289} (1987) 1;\hfill
G.C. Branco and M.N. Rebelo, Phys.\ Lett.\ {\bf B160} (1985) 117}

\preref\Weinberg{%
S. Weinberg, Phys.\ Rev.\ {\bf D42} (1990) 860}

\preref\Mendez{%
A. Mendez and A. Pomarol; Phys.\ Lett.\ {\bf B272} (1991)
 313}

\preref\BSP{%
W. Bernreuther, T. Schr\"oder, and T.N. Pham, Phys.\ Lett.\ {\bf B279}
(1992) 389}

\preref\GGone{%
B. Grzadkowski and J. Gunion, Phys.\ Lett.\ {\bf B289} (1992) 440}

\preref\Peskin{%
C.R. Schmidt and M.E. Peskin, Phys.\ Rev.\ Lett.\ {\bf 69} (1992) 410}

\preref\FMK{%
C.D. Froggatt, R.G. Moorehouse,
and I.G. Knowles, Nucl.\ Phys.\ {\bf B386} (1992) 63}

\preref\Soni{%
D. Atwood and A. Soni, Phys.\ Rev.\ {\bf D45} (1992) 2405;
S. Bar-Shalom et al. SLAC preprint SLAC-PUB-6765 (1995)}

\preref\GGtwo{%
B. Grzadkowski and J. Gunion, Phys.\ Lett.\ {\bf B294} (1992) 361}

\preref\BBone{%
W. Bernreuther and A. Brandenburg, Phys.\ Lett.\
{\bf B314} (1993) 104}

\preref\Changone{%
D. Chang and W.-Y. Keung: Phys.\ Lett.\ {\bf B305} (1993)
261}
\preref\Changtwo{%
D. Chang, W.-Y. Keung, and I. Phillips,
Phys.\ Rev.\ {\bf D48} (1993) 3225}

\preref\BBtwo{%
W. Bernreuther and A. Brandenburg,
Phys.\ Rev.\ {\bf D49} (1994) 4481}

\preref\BO{%
W. Bernreuther and P. Overmann, Z. Phys.\ {\bf C61} (1994) 599}

\preref\Pil{%
A. Pilaftsis and M. Nowakowski,  Int. J. Mod. Phys.
{\bf A9} (1994) 1097; ibid.\ {\bf A9} (1994) 5849 (E)}

\preref\Maone{%
X.G. He, J.P. Ma, and B.H.J. McKellar, Mod. Phys.\ Lett.\ {\bf A9}
(1994) 205; Phys.\ Rev.\ {\bf D49} (1994) 4548}

\preref\Grone{%
B. Grzadkowski, Phys.\ Lett.\ {\bf B338} (1994) 71}

\preref\Matwo{%
J.P. Ma and
B.H.J. McKellar, Melbourne preprint UM-P-94/50 (1994)}
\preref\Ginz{%
I.F. Ginzburg et al., Nucl.\ Instrum.\ Meth.\ {\bf 205} (1983) 47;
 ibid. {\bf 219} (1984) 5;
V.I. Telnov, Nucl.\ Instr.\ Meth.\ {\bf 294} (1990) 72}

\preref\Zerwas{%
{\it $e^+ e^-$ Collisions at 500 GeV: The Physics Potential},
ed. by P.M. Zerwas,
DESY publication DESY 92-123A,B, Hamburg 1992, DESY 93-123C (1993);
{\it Physics and Experiments with Linear Colliders},
ed.\ by R. Orava, P. Eerola, and
M. Nordberg (World Scientific, Singapore), Vols.\ I, II (1992)}

\preref\Zerwastwo{%
M. Kr\"amer, J. K\"uhn, M. Stong, and P. Zerwas,
Z. Phys.\ {\bf C64} (1994) 21}

\preref\Osland{%
A. Skjold and P. Osland, Phys.\ Lett.\ {\bf B329} (1994) 305}

\preref\Seone{%
T. Arens, U.D.J. Giesler, and L.M. Sehgal,
Phys.\ Lett.\ {\bf B339} (1994) 127}

\preref\Setwo{%
T. Arens and L.M. Sehgal, Aachen preprint PITHA-94-37 (1994)}
\preref\BMM{%
W. Bernreuther, J.P. Ma, and B.H.J. McKellar,
 Phys.\ Rev.\ {\bf D51} (1995) 2475}

\preref\Hollik{%
F. Cornet, W. Hollik, and W. M\"osle, Nucl.\ Phys.\ {\bf B428} (1994) 61}

\preref\BNOS{%
W. Bernreuther, O. Nachtmann, P. Overmann, T. Schr\"oder,
Nucl.\ Phys.\ {\bf B388} (1992) 53; ibid.\ {\bf B406} (1993) 516 (E)}

\preref\BraMa{%
J.P. Ma, A. Brandenburg, Z. Phys.\ {\bf C56} (1992) 97}

\preref\Kuehn{%
J.H. K\"uhn, E. Mirkes, and J. Steegborn, Z. Phys.\ {\bf C57} (1993) 615}


$\null$
\vskip -3. cm

\section{Introduction}
\tagsection\IntroSection
CP-violating interactions beyond the Kobayashi-Maskawa mechanism
and their high-energy phenomenology have been investigated rather intensely
in recent years.
This was (and is) motivated to some extent
by proposals of efficient non-standard model
scenarios for generating the cosmological baryon asymmetry at
the electroweak phase transition (for a review, see [\use\Cohen].)
An extended Higgs boson sector -- as predicted by many
extensions of the Standard Model (SM) -- provides such interactions
in a natural way [\use\Lee,\use\Weinberg].  In the case
of Higgs sector CP violation one expects, in particular, a
 spectrum of neutral Higgs particles with indefinite  CP parity.
This can be traced through large Yukawa
 couplings to heavy fermions, notably to top quarks
[\use\BSP,\use\GGone,\use\Peskin,\use\Soni,\use\GGtwo,\use\BBone,\use\Changone
 ,\use\Changtwo,\use\BBtwo,\use\BO,\use\Pil,\use\Maone,\use\Grone,\use\Matwo].

A ``Compton collider'' [\use\Ginz] which is considered as an option
 in the context of the present
discussion of high-energy  $e^+ e^-$ linear colliders,
 would provide, among other things,
 an interesting possibility to produce neutral Higgs bosons and
 to study their quantum numbers [\use\Zerwas,\use\Zerwastwo].
 In this letter we investigate, in the framework of two-Higgs doublet
extensions of the SM with explicit CP violation in the Higgs potential
[\use\Weinberg,\use\BSP,\use\Mendez,\use\FMK],
CP-violating effects in unpolarized {\footnote{\hskip -0.15cm$^\dagger$} \
CP asymmetries
for polarized
$\gamma \gamma \to\varphi\to t\bar{t}$ were investigated in [\use\GGtwo].
They apply if the
polarizations of both photons are adjustable.}
  $\gamma \gamma \rightarrow t \bar{t} $
which includes, for neutral Higgs boson masses above the
$t \bar{t}$ threshold, resonant $\varphi$ production and the subsequent
decay of $\varphi \rightarrow t \bar{t}$. If $\varphi$ is not a CP eigenstate,
a CP-violating spin-spin correlation is induced in the decay
$\varphi \rightarrow t \bar{t}$ (and in the decays into other
fermions, respectively) already at the Born level
 which can be as large as 0.5, as pointed out in [\use\BBone]
(see also
[\use\Changtwo,\use\Maone], and for absorptive effects see
[\use\Maone,\use\Grone,\use\Osland,\use\Seone,\use\Setwo]).
However,
the narrow-width approximation does not apply for a Higgs boson with
mass above the $t\bar{t}$ threshold and interference with the non-resonant
$\gamma \gamma \rightarrow t \bar{t}$ amplitude
decreases this spin-spin correlation
significantly, as shown below. As we wish to consider also the case of light
Higgs bosons $\varphi$ below the $t \bar{t}$ threshold we have computed, for
the above models, the complete set of CP-nonconserving contributions
to $\gamma \gamma \rightarrow t \bar{t}$ in one-loop approximation.
Apart from the above spin-spin correlation we determine also a
polarization asymmetry which projects onto absorptive CP effects.
In addition we propose and calculate
three asymmetries with which this polarization asymmetry and the
above-mentioned spin-spin correlation can be traced
efficiently in  semileptonic $t \bar{t}$ decays.

\vfil\eject
\section{CP violation in $\gamma \gamma \rightarrow t \bar{t} $}
\tagsection\kinematicssection

We first discuss signatures of CP violation in the reaction

 $$
\gamma (p_1)+\gamma (p_2) \rightarrow t(k_1) +\bar t(k_2),
\eqn\reac
$$
\noindent
where the momenta are defined in the photon-photon
CM frame. We consider only unpolarized photon beams. The initial
state of (\reac) then forms a CP eigenstate.
The process may  be described by the density matrix:

$$
 R_{\alpha \alpha ',\beta \beta '}({\bf p, k}) = \hbox{$\sum$}^{'}
 \langle t(k_1,\alpha ')\bar t(k_2,\beta ') \vert {\cal T}
  \vert \gamma(p_1)\gamma(p_2) \rangle^*
  \langle t(k_1,\alpha )\bar t(k_2,\beta ) \vert{\cal T}
  \vert \gamma(p_1)\gamma(p_2) \rangle
\eqn\density
$$
\noindent
where the $\sum^{'}$ denotes averaging over the $\gamma \gamma$
polarizations,
$\alpha, \alpha ', \beta, \beta '$ are spin indices, and ${\bf p}={\bf p}_1,
{\bf k}={\bf k}_1$.
Note that $R$ is an even function of the three-momentum
${\bf p}$ due to Bose symmetry of the
two-photon state. The squared matrix  element for the process
$\gamma \gamma \rightarrow t \bar{t} \rightarrow X $ is then given,
in the narrow-width approximation for the $t$ quark, by
tr$(R \rho_t \rho_{\bar t})$, where  $\rho_t$ and $\rho_{\bar t}$ are the
decay density matrices for polarized $t$ and $\bar t$ decay, respectively.

The general structure
of CP-violating contributions to the production density matrix
$R$ can be determined easily. For unpolarized photons one
finds that CP-violating dispersive contributions must be of the form
[\use\BBone,\use\BBtwo]

$$\eqalign{
&  \hat{\bf k}\cdot ({\bf s}_+\times{\bf s}_-) h_e(y), \cr
&
\hat{\bf p}\cdot ({\bf s}_+\times{\bf s}_-) h_o(y),
}
\eqn\dispobs
$$
\noindent
where ${\bf s}_+$, ${\bf s}_-$ are the spin operators of
$t$ and $\bar{t}$, respectively, and
$h_e(y)$, $h_o(y)$ are even and odd functions
 of the cosine of the scattering angle,
$y = \hat{\bf p}\cdot \hat{\bf k}$.
Note that QCD- or QED-induced absorptive parts of the scattering amplitude
of the process (\reac) cannot induce $t \bar{t}$ spin-spin correlations:
they generate a polarization of $t$ and $\bar t$ normal to the scattering
plane of the reaction (\reac) [\use\BBone,\use\BMM].
CP-violating absorptive contributions to $R$ are of the form
$$
\eqalign{
& \hat{\bf k}\cdot ({\bf s}_+-{\bf
s}_-)f_e(y),\cr
&\hat{\bf p}\cdot ({\bf s}_+-{\bf s}_-)f_o(y),
}
\eqn\absobs
$$
\noindent
where $f_e$ and $f_o$ are even and odd functions of $y$, respectively.

We now discuss the salient features of neutral Higgs sector CP violation.
For definiteness we consider  two-doublet extensions of the SM with
explicit CP violation in the Yukawa couplings (which leads to the
Kobayashi-Maskawa phase) and in the Higgs potential
[\use\Weinberg,\use\BSP,\use\Mendez,\use\FMK]. As a consequence
the three physical neutral Higgs boson states $\varphi_j, j=1,2,3$ are
in general states
with indefinite CP parity; i.e., they couple both to scalar and
pseudoscalar quark and lepton currents with strength
$ a_{jf} m_f/v$ and $\tilde{a}_{jf} m_f/v$, respectively, where $m_f$ is
the fermion mass and $v\simeq 246$ GeV. For the top quark we have

$$
a_{jt}=d_{2j}/\sin \beta,\ \ \ \tilde{a}_{jt}=-d_{3j}\cot \beta,
\eqn\coupl
$$
\noindent
where  $\tan \beta=v_2/v_1$ is the ratio of the moduli of the
 vacuum expectation
values of the two doublets, and $d_{2j},\ d_{3j}$ are the matrix elements
of a $3\times 3$ orthogonal matrix which describes the mixing
of the neutral  Higgs states of definite CP
parity.
Only the CP=+1 components of the mass eigenstates $\varphi_j$ couple to
the $W$ and charged Higgs bosons at the tree level.
(For notation and details, see [\use\BSP]).
\par
CP violation requires that the neutral Higgs bosons are not mass-degenerate.
In the following we assume that the masses of $\varphi_{2,3}$ are much larger
than the mass of $\varphi_1$ and  also larger than
the photon-photon CM energy. Then the
effect of $\varphi_{2,3}$ on the
quantities discussed below is negligible.

The Born amplitude for the reaction (\reac)
and the contributions from
$\varphi$ exchange at one loop are depicted in Fig.1. (Note that there is
no CP-violating
contribution from the Kobayashi-Maskawa phase to
this order in pertubation theory.)
Figs.1b -- 1e represent CP-violating contributions which are proportional
to the coupling $a_{1t}\tilde{a}_{1t}$.
A remark concerning Fig.1e is in order:
the CP-violating $\varphi_j$ exchange
contributions to the self energy of the top quark are of
the form $\Sigma_{CP}(p^2) = m_tf(p^2)\gamma_5$.
The function $f(p^2)$ is
actually finite
if one sums over all $\varphi_j$ and takes into account the
orthogonality properties of the mixing matrix $d_{ij}$.
For the sake of simplicity and in the spirit of the previous discussion,
we keep only the contribution from $\varphi_1$. We use an on-shell
definition of the top mass.
This induces a counterterm with Lorentz structure
$m_tf(m^2_t)\gamma_5$,
which has to be taken into account in Fig.1e. \par\noindent
If the mass of $\varphi$ is
close to the $\gamma\gamma$ CM energy,
the contributions
Fig.1f--1h become resonant.
(Fig.1f represents four amplitudes: two CP-conserving
ones with couplings $a_{1t}^2$ and $\tilde{a}_{1t}^2$, respectively, and two
CP-violating ones with couplings $a_{1t}\tilde{a}_{1t}$. Likewise,
Figs.1g,h represent two amplitudes where $\varphi$ couples to the scalar
top current and two amplitudes with  $\varphi$ coupling
to the pseudoscalar current.)
Even in the resonant case
 interference of these terms with the Born amplitude
is non-negligible because of the finite width
of $\varphi$ [\use\BBone]. We compute this width in the two-doublet model
by summing the partial widths for $\varphi \rightarrow W^+W^-,ZZ,t\bar{t}$,
assuming that the charged Higgs is heavy. (For definiteness we take in
the following $m_{H^\pm}=500$ GeV.)
\par\noindent
The CP-conserving part of the density matrix ~(\density) is determined from
the squared Born amplitude Fig.1a, the interference of Fig.1a with the CP-even
amplitudes of Fig.1f,g,h and the square of the sum of Figs.1f,g,h. The CP-odd
part of $R$ results from the interference of the Born diagram with the CP-odd
terms from Figs.1b -- 1h and the interference of the CP-even and -odd
amplitudes of Figs.1f -- 1h.
For Higgs boson masses of the order of $100$ GeV or larger neutral Higgs
sector CP violation is so far not very stringently constrained
by low-energy phenomenology, which includes the experimental
upper bounds on the electric dipole moments of the electron and neutron.
Below and in the next section we evaluate CP-asymmetries for the following
set of parameters:
$$
d_{i1} = 1/\sqrt 3,\ \  i=1,2,3; \ \ \ \ \ \ \tan \beta = 0.5
\eqn\parone
$$
and
$$ d_{i1} = 1/\sqrt 3,\ \ i=1,2,3; \ \ \ \ \ \ \tan \beta = 1.0
\eqn\partwo
$$
\noindent
(Values of $\tan \beta$ as small as
$0.5 $ can be accommodated by  phenomenology [\use\Hollik].)
Set (\parone) is used to
exhibit the maximal order of magnitude of the asymmetries below
within the two-doublet models.

For illustrative purposes we have plotted in Figs.2,3 two basic CP-odd
 correlations at the parton level, using the above parameter set (\partwo),
$m_t = 175$ GeV, and $m_{\varphi} = 400$ GeV.
(The correlations are normalized such that $\langle 1\rangle = 1$.)
Figs.2,3 show the longitudinal polarization asymmetry
$\langle\hat{\bf k}\cdot ({\bf s}_+-{\bf s}_-)\rangle$
and the
spin-spin correlation  $\langle\hat{\bf k}\cdot ({\bf s}_+\times{\bf s}_-)
\rangle$,
respectively, as a function of the photon-photon CM energy
$\sqrt{s_{\gamma\gamma}}$.
The  spin-spin correlation is largest
slightly below $\sqrt{s_{\gamma\gamma}}=m_{\varphi}$,
but then changes sign due to the
interference of the various amplitudes. Only in the unrealistic limit
$\Gamma_{\varphi}/m_{\varphi} \rightarrow 0$ the naive Born level result
mentioned in Sect.1 would be recovered [\use\BBone]. For $\tan\beta=0.5$
the correlations shown in Figs. 2,3 increase roughly by a factor of two.

\section{CP asymmetries for semileptonic $t \bar{t}$ decays}
\tagsection\asymmetrysection

In this section we evaluate a few  asymmetries which efficiently trace
the ``basic'' CP-odd quantities ~(\dispobs) and ~(\absobs) in the final
states into which $t$ and $\bar{t}$ decay.
\noindent
We define a sample ${\cal A}$ as consisting of $t\bar{t}$ pairs where the
$t$ decays semileptonically  and the $\bar{t}$
decays hadronically:

$$
\eqalign{
t\ &\to\
\ell^++\nu_{\ell} + b \cr
\bar{t}\ &\to\  W^-+\bar{b}
\to q+\bar{q}'+\bar{b}
,} \eqn\decays
$$
\noindent
The sample  ${\cal B}$ is  defined by the charge
conjugated decay channels of the $t\bar{t}$ pairs with respect to sample
${\cal A}$.

We concentrate here on the above decay modes, since they
are especially suited for our CP studies:
In each event we have one lepton which is known to be a good analyser
of the top spin %
{\footnote{$^\ddagger$} \
We use the distributions for the decays of polarized $t \rightarrow
\ell, t \rightarrow W,\ldots$ in the form as given in
[\use\BNOS,\use\BraMa].}, and in the same event we may
reconstruct the momentum of the other top quark from its hadronic
decay products.
Note that in the $e^+e^-$ (or $e^-e^-$)  laboratory system, the top
quark pairs are in general not produced back to back, since the
backscattered photons carry different energy fractions which cannot
be determined event by event. Thus only
the rest system of one of the top quarks may be reconstructed
unambiguously in
each of the above samples. Purely hadronic
decays which would allow in principle for a reconstruction of both top
and antitop momenta
might also be considered. Because flavour-tagging for $W\to q\bar{q}'$ is
probably inefficient,
the reconstructed directions of flight
 of the $W^+ (W^-)$ must then act as
analysers of the $t (\bar{t})$ spin.
 However, they have a lower analyser quality
than the charged lepton.
The channels where both $t$ and $\bar{t}$ decay semileptonically
could also be used for CP studies. However, apart from
having a smaller branching ratio the momentum (direction) of the
top cannot be reconstructed for these modes. This leads,
generally speaking, to a decrease in
sensitivity to the effects which we are after.

Therefore we confine ourselves to the event samples ${\cal A}$
and  ${\cal B}$. First we consider the observables:

$$
\eqalign{
{\cal O}_1\ &=\ ({\bf q}^L_{\ell +}\times \hat{{\bf q}}^*_{W-})
\cdot \hat{\bf{k}}_{\bar{t}}^L\ \ \ \ \ \ \ \
{\rm for \ sample}\  {\cal A}, \cr
\bar{{\cal O}}_1\ &=\ ({\bf q}^L_{\ell -}\times \hat{{\bf q}}^*_{W+})
\cdot \hat{{\bf k}}_{t}^L\ \ \ \ \ \ \ \ {\rm for \ sample}\  {\cal B}
.}
\eqn\obsone
$$
\noindent
Here, the upper index ``$L$'' refers to the overall laboratory system and
the asterisk denotes the $\bar{t}$($t$) rest system.
A CP-odd ``asymmetry'' may be defined through the sum

$$ {\cal A}_1\ =\
\langle {\cal O}_1 \rangle_{\cal A}+
\langle \bar{{\cal O}}_1 \rangle_{\cal B}.
\eqn\asymmone
$$
\noindent
${\cal A}_1$ traces the CP-odd spin-spin correlations defined
in ~(\dispobs).

\noindent We also define, for the same samples, the observables

$$\eqalign{
{\cal O}_2 \ &=\  E_{\ell+}^L,\cr
{\bar {\cal O}}_2\ &=\ E_{\ell-}^L,
}
\eqn\obstwo
$$

$$
\eqalign{
{\cal O}_3\ &=\ {\bf q}^L_{\ell +}\cdot \hat{{\bf k}}^L_{\bar{t}},\cr
\bar{{\cal O}}_3\ &=\ {\bf q}^L_{\ell -}\cdot \hat{{\bf k}}^L_t
}
\eqn\obsthree
$$

\noindent and the CP asymmetries

$$
{\cal A}_2 \ =\   \langle{\cal O}_2 \rangle_{\cal A}-
 \langle {\bar {\cal O}}_2 \rangle_{\cal B},
\eqn\asymmtwo
$$

$$
{\cal A}_3 \ =\
\langle {\cal O}_3 \rangle_{\cal A}
-\langle \bar{{\cal O}}_3 \rangle_{\cal B}.
\eqn\asymmthree
$$

\noindent The quantities (\asymmtwo) and (\asymmthree) probe
CP violation generated by the asymmetries in the $t$ and $\bar{t}$
polarizations in the production plane  (\absobs).

To  calculate  the above correlations, we have to convolute
the ``partonic'' differential cross section for $\gamma\gamma\to t\bar{t}$
and $t\bar{t} \to {\cal A}$ or ${\cal B}$
with the distribution function for the backscattered laser photons.
We neglect the transverse momenta of the Compton photons, which is a very
good approximation [\use\Ginz].
The differential distributions are
of the form tr$(R\rho_t \rho_{\bar{t}})$, where
the decay distributions  $\rho_t$ and $\rho_{\bar{t}}$ are taken
from [\use\BNOS,\use\BraMa].  We then obtain the asymmetries
(\asymmone), (\asymmtwo), (\asymmthree) as integrals over some elements
of the density matrix $R$
defined in (\density). For example,

$$\eqalign{
{\cal A}_2 \ &=\ {4\over 3}{1+2\omega+3\omega^2\over 2+4\omega}
{1\over 2s}{1\over \sigma_0} \int_0^{xmax} {dx_1\over x_1} N(x_1)
\int_0^{xmax} {dx_2\over x_2} N(x_2) \int_{-1}^1 dy {\beta\over 16\pi}
\cr &\times\
\beta E{x_1+x_2\over 2\sqrt{x_1 x_2}}(yb_1^{CP}+b_2^{CP})
.}
\eqn\resultE
$$

\noindent
Here $N(x)$ is the normalized distribution of photons with energy
fraction $x=E_{\gamma}/E_{beam}$,
 $\omega=m_W^2/m_t^2$, $s$ denotes the squared $e^+e^-$ CM energy,
and $\sigma_0$ is the effective cross section for
$\gamma\gamma\to t\bar{t}$, i.e., the $\gamma\gamma\to t\bar{t}$ cross
section folded with the photon distribution functions.
 Further,
$E$ is the energy and $\beta=\sqrt{1-m_t^2/E^2}$ is the velocity of the
top quark in the $\gamma\gamma$ CM system.
The distributions turn  out to be
essentially flat as functions of the cosine $y$ of the scattering
angle in the $\gamma\gamma$  CM  frame.
Therefore we do not apply a cut in this variable. The functions $b_{1,2}^{CP}$
are the CP-violating absorptive contributions to the density matrix $R$.

For $N(x)$  we take  the leading order
result, as given e.g. in [\use\Ginz,\use\Kuehn]. The maximal energy
fraction of a photon
$x_{max}$
is determined by the laser energy,

$$
\eqalign{
x_{max} \ &=\ {z\over 1+z}\ \ \ \ \ \ {\rm with} \cr
z\ &=\ {4E_{beam}E_{laser}\over m_e^2}.
}
\eqn\Laser
$$

\noindent We choose the laser energy
$E_{laser}$  for a given beam energy $E_{beam}$ such that $z$ reaches
its maximal value, $z_{max}=2(1+\sqrt{2})$
, which is determined
by the threshold of the undesirable production of $e^+e^-$ pairs
through annihilation of a backscattered photon with a laser photon.
Thus we have $x_{max}\approx 0.8284 $.
\par
Formulas analogous to (\resultE) hold also for the other asymmetries.
In order to estimate the statistical sensitivity of
${\cal A}_i\ (i=1,2,3)$, we have computed the ratios
${\cal A}_i/\Delta {\cal A}_i$, where
$\Delta{\cal A}_i
\equiv \Delta {\cal O}_i\approx \sqrt{\langle {\cal O}_i^2\rangle_{{\cal A}}}
\approx \sqrt{\langle\bar{{\cal O}}_i^2\rangle_{{\cal B}}}$. The
corresponding signal-to-noise ratios $S_i$ are given by

$$
S_i={|{\cal A}_i|\over \Delta {\cal A}_i}\times \sqrt{N_{event}},
\eqn\statsig
$$

\noindent where

$$
N_{event}=R_{\cal A,B}{\cal L} \sigma_{0}
\eqn\events
$$

\noindent
with the integrated luminosity $\cal{L}$ of the Compton collider and the
branching ratios $R_{\cal A,B}$ for the decay modes
${\cal A},{\cal B}$.
\par\noindent
Figs. 4-6 show the three ratios ${\cal A}_i/\Delta {\cal A}_i$, $i=1,2,3$
for different Higgs masses $m_{\varphi}$ at
a fixed $e^+e^-$ collider energy $\sqrt{s}=500 $ GeV.
In each figure, the dashed line corresponds
to the parameter set (\parone) and the full line to the parameter set
(\partwo).
The asymmetries exhibit extrema when the Higgs
mass is close to the $t\bar{t}$ threshold due to large interference
effects, and additional extrema when the Higgs mass is close to the maximal
$\gamma\gamma$ energy.  Even for a Higgs with mass above the maximal
$\gamma\gamma$ energy, there remains a significant interference effect
that might be detectable.
\par
For the most sensitive of our asymmetries, ${\cal A}_3$,
we also determined, for a given Higgs mass, the collider
energy that maximizes the signal-to-noise ratio. The results
are listed in Table 1 for the parameter set (\partwo). In brackets
we also give
the numbers for set (\parone).
We use only semileptonic top decays into electrons and muons,
 i.e., $R_{\cal A,B}=4/27$. For example,
at $m_{\varphi}=400$ GeV, one gets the largest sensitivity for
a collider energy of $490$ GeV.
For an integrated luminosity of
100~fb$^{-1}$ we then find a statistical significance
$S_3$ of 3.1(4.9). This indicates
the exciting possibility of probing  Higgs sector CP-violation.


\section{Conclusions}

A high luminosity ``Compton collider'' would offer an interesting
opportunity
to study in detail the neutral Higgs sector and in particular
CP violation beyond the Kobayashi-Maskawa mechanism which is
possible in multi-Higgs extensions of the standard model.
In this paper we have computed for
two-doublet models the complete CP-violating contributions
to unpolarized $\gamma\gamma\to t\bar{t}$ in one-loop approximation.
The dispersive contributions give rise to CP odd spin-spin correlations
which can reach values of the order of ten percent. Longitudinal
polarization asymmetries of similar
magnitude are induced by CP-violating absorptive
contributions. We have further proposed and calculated three
CP asymmetries for semileptonic $t\bar{t}$ decays which efficiently
trace the CP odd effects considered. We find that for quite a broad window
of Higgs masses, a high luminosity Compton collider operating
at an $e^+e^-$ CM energy of $\sqrt{s}=500-600$ GeV would allow for the
possiblity
to probe CP-violating effects from an extended Higgs sector.

\vskip .2 cm

\noindent{\bf Acknowledgments.}

\vskip .1 cm

Two of us (H.A. and A.B.) would like to thank the SLAC Theory Group for
the warm hospitality extended to them during their stay where part of this
work was done.

\listrefs
\centerline{\bf Figure Captions}
\vskip .3 cm
\noindent{\bf Figure~1:}
Feynman diagrams for $\gamma\gamma\to t\bar{t}$. In (a) the Born
diagram is shown, (b) to (h) depict the relevant one-loop
 $\varphi$ exchange diagrams. In (g) and (h), we show the contributions
due to $W$ boson, charged Higgs $H$, Goldstone boson $G$ and ghost $\eta$
propagation around the loop. Diagrams with crossed photon lines are
not shown.\par
\vskip .1 cm
\noindent{\bf Figure~2:}
Longitudinal polarization asymmetry $\langle \hat{\bf{k}}\cdot (\bf{s}_+
-\bf{s}_-)\rangle$ as a function of the photon-photon CM energy for
$m_{\varphi}=400$ GeV, $m_t=175$ GeV and parameter set (\partwo).
\par
\vskip .1 cm
\noindent{\bf Figure~3:}
Spin-spin correlation  $\langle \hat{\bf{k}}\cdot (\bf{s}_+
\times \bf{s}_-)\rangle$ as a function of the photon-photon CM energy for
the same choice of parameters as in Fig. 2.
\par
\vskip .1 cm
\noindent{\bf Figure~4:}
The quantity  ${\cal A}_1/\Delta {\cal A}_1$ for different Higgs masses
$m_{\varphi}$ at $\sqrt{s}=500$ GeV.
The dashed line corresponds to parameter set (\parone),
the full line to parameter set (\partwo).
\par
\vskip .1 cm
\noindent{\bf Figure~5:}
Same as Fig.4, but for ${\cal A}_2/\Delta {\cal A}_2$.
\par
\vskip .1 cm
\noindent{\bf Figure~6:}
Same as Fig.4, but for ${\cal A}_3/\Delta {\cal A}_3$.
\vfil\eject

\centerline{\bf Table Caption}
\vskip .3 cm
\noindent{\bf Table~1:}
Optimal collider energies $\sqrt{s_{opt}}$, number of events
$N_{event}$ for samples ${\cal A,B}$
(cf. 3.1) and statistical significance $S_3$ of the asymmetry
$A_3$ for some Higgs masses $m_{\varphi}$. In computing $N_{event}$
and $S_3$ only semileptonic $t$ decays into $e$ and $\mu $ are
taken into account; i.e., we take $R_{\cal A,B}=4/27$.
To obtain
the optimal collider energy, the parameter set
(\partwo) is used. The numbers for parameter set (\parone) at
the same energies are
given in brackets.

\vfil\eject
%
%
\centerline{\bf{Table 1}}
\vskip 0.3cm
$$
{\offinterlineskip \tabskip=0pt
\vbox{
\halign{ \strut
         \vrule#&
\hfill # \hfill  &
         \vrule#&
 \hfill # \hfill &
\vrule#&
 \hfill # \hfill  &
\vrule#&
\hfill # \hfill &
\vrule#
\cr
\noalign{\hrule}
 & $m_{\varphi}\left[{\rm GeV}\right]$
&& $\sqrt{s_{opt}}\left[{\rm GeV}\right]$
&& $N_{event}/\left(
{\cal L}/(100\ {\rm fb}^{-1})\right)$ &&
$S_3/\sqrt{{\cal L}/(100\
 {\rm fb}^{-1})}$ &  \cr
\noalign{\hrule}
& 100 && 690 && $4.46(4.28)\times 10^3$ &&  $1.2(2.8)$ & \cr
& 150 && 690 && $4.46(4.27)\times 10^3$ &&  $1.2(2.8)$ & \cr
& 200 && 690 && $4.44(4.24)\times 10^3$ &&  $1.2(3.0)$ & \cr
& 250 && 670 && $4.15(3.88)\times 10^3$ &&  $1.4(3.3)$ & \cr
& 300 && 630 && $3.42(3.13)\times 10^3$ &&  $1.7(4.2)$ & \cr
& 325 && 590 && $2.64(2.31)\times 10^3$ &&  $2.1(5.2)$ & \cr
& 350 && 540 && $1.60(1.42)\times 10^3$ &&  $3.0(7.3)$ & \cr
& 375 && 620 && $3.27(3.17)\times 10^3$ &&  $2.7(5.9)$ & \cr
& 400 && 490 && $0.88(1.11)\times 10^3$ &&  $3.1(4.9)$ & \cr
& 425 && 520 && $1.48(1.70)\times 10^3$ &&  $3.1(5.1)$ & \cr
& 450 && 550 && $2.13(2.31)\times 10^3$ &&  $2.8(4.7)$ & \cr
& 475 && 580 && $2.74(2.92)\times 10^3$ &&  $2.6(4.2)$ & \cr
& 500 && 610 && $3.32(3.44)\times 10^3$ &&  $2.3(3.8)$ & \cr
\noalign{\hrule} }}
}
$$

\end